\documentclass[%
 preprint,
 amsmath,amssymb,
 aip,
 cha,
 numerical
]{revtex4-1}

\usepackage{graphicx}% Include figure files
\usepackage{dcolumn}% Align table columns on decimal point
\usepackage{bm}% bold math
\usepackage{hyperref}% add hypertext capabilities
\usepackage{color}

\begin{document}

%\preprint{APS/123-QED}

\title{Ott-Antonsen attractiveness for parameter-dependent oscillatory networks}

\author{Bastian Pietras} 
\email{b.pietras@vu.nl}
\affiliation{%
 MOVE Research Institute Amsterdam \& Institute for Brain and Behavior Amsterdam, Faculty of Behavioural and Movement Sciences, Vrije Universiteit Amsterdam, van der
Boechorststraat 9, Amsterdam 1081 BT, The Netherlands}
\affiliation{Department of Physics, Lancaster University, Lancaster LA1 4YB United Kingdom}

\author{Andreas Daffertshofer}
\email{a.daffertshofer@vu.nl}
\affiliation{%
 MOVE Research Institute Amsterdam \& Institute for Brain and Behavior Amsterdam, Faculty of Behavioural and Movement Sciences, Vrije Universiteit Amsterdam, van der
Boechorststraat 9, Amsterdam 1081 BT, The Netherlands}

%\affiliation{%
% MOVE Research Institute Amsterdam \& Institute for Brain and Behavior Amsterdam, Faculty of Behavioural and Movement Sciences, Vrije Universiteit Amsterdam, van der
%Boechorststraat 9, Amsterdam 1081 BT, The Netherlands}


%\affiliation{
%\affiliation{
%\affiliation{%


\date{\today}% It is always \today, today,

\begin{abstract}
The Ott-Antonsen (OA) ansatz [Chaos 18, 037113 (2008), Chaos 19, 023117 (2009)] has been widely used to describe large systems of coupled phase oscillators. If the coupling is sinusoidal and if the phase dynamics does not depend on the specific oscillator, then the macroscopic behavior of the systems can be fully described by a low-dimensional dynamics. Does the corresponding manifold remain attractive when introducing an intrinsic dependence between an oscillator's phase and its dynamics by additional, oscillator specific parameters? To answer this we extended the OA ansatz and proved that parameter-dependent oscillatory systems converge to the OA manifold given certain conditions. Our proof confirms recent numerical findings that already hinted at this convergence. Furthermore we offer a thorough mathematical underpinning for networks of so-called theta neurons, where the OA ansatz has just been applied. 
In a final step we extend our proof by allowing for time-dependent and multi-dimensional parameters as well as for network topologies other than global coupling. This renders the OA ansatz an excellent starting point for the analysis of a broad class of realistic settings.
\end{abstract}
\maketitle

\begin{quotation}
Coupled phase oscillators are being widely used to describe synchronization phenomena.
The study of their collective dynamics has experienced a major breakthrough by the results by Ott and Antonsen \cite{OttAntonsen2008, OttAntonsen2009, OttAntonsenComment}.
The asymptotic behavior of the mean field of infinitely many coupled oscillators can be cast into a reduced, low-dimensional system of ordinary differential equations. The evolution is hence captured by the so-called Ott-Antonsen (OA) manifold.

Very recently, the OA ansatz has been applied to networks of theta neurons, see, e.g., Refs.~\onlinecite{luke2013complete, so2014networks, laing2014derivation,MontbrioPazoRoxin2015,Laing2016, Byrne2016, Coombes2016}.
A particular property of coupled, inhomogeneous theta neurons is that both the phase of a single neuron as well as its dynamics depend on a parameter, which establishes an intrinsic relation between them.
While numerical results suggest the attractiveness of the OA manifold in the presence of such a parameter dependence, it has as to yet not been proven whether the dynamics really converges to it.
For a certain class of parameter dependencies we here extend the existing theory of the OA ansatz and show that the OA manifold continues to asymptotically attract the mean field dynamics.

Parameter-dependent systems and their description through the OA ansatz have been considered by, e.g.,
 Strogatz and co-workers\cite{MarvelMirolloStrogatz2009}, Wagemaker and co-workers\cite{Wagemakers2014}, and So and Barreto\cite{So2011}.
There, parameters seemingly did not yield a correlation between an oscillator's phase and its dynamics but a rigorous proof for this is still missing.
We explicitly address this last point.
In particular, we prove a conjecture later formulated by Montbri\'o and co-workers\cite{MontbrioPazoRoxin2015} on the attractiveness of the OA manifold for parameter-dependent systems.
The case of parameters serving as mere auxiliary variables readily follows from our result -- we will refer to this as ``weak" parameter-dependence\cite{Note1}.
%\footnote{Parameter-dependent systems comprise a wide class of systems, from which we here only choose a single family. This family represents a rather weak parameter-dependent system. However, we refrain from this notion since weak parameter-dependence would imply that parameter changes have little to no considerable effect. Here, the original proof by Ott and Antonsen has to be changed, such that the parameter effect can be strong. We use the attribute ``weak'' to highlight that specific oscillator does not depend on the additional parameter but its mean field dynamics only.}.
By showing that a network of theta neurons can be treated as a parameter-dependent oscillatory system, our result establishes an immediate link to networks of quadratic integrate-and-fire (QIF) neurons:
That is, the so-called Lorentzian ansatz as an equivalent approach to the OA ansatz is analytically substantiated.
By this we may exert an important impact in mathematical neuroscience.

Finally, we extend the parameter-dependence for more general classes of networks.
First, we address non-autonomous systems and show that our proof can be applied to time-varying parameters.
An important example here is a biologically realistic approach to oscillatory systems proposed by Winfree\cite{Winfree1967}.
Second, we include multiple distributed parameters illustrated by coupled limit-cycle oscillators with shear.
Third, we apply our proof to networks with different coupling topologies including non-local coupling by using an heterogeneous mean field approach.\\
\end{quotation}

\section{Introduction}
The Kuramoto model can be considered the most seminal description of globally coupled networks of phase oscillators.
It has been investigated in great detail but its various extensions still make it {\em{the}} model-to-work-with when it comes to the study of network dynamics \cite{Acebron_Review2005,Rodrigues2016}.
We adopt the notion of Montbri\'o, Paz\'o, and Roxin\cite{MontbrioPazoRoxin2015} and write the Kuramoto-like model as
\newcounter{eq_dummy}\setcounter{eq_dummy}{1}%
\begin{equation}
\label{single1}
\dot{\theta_j} = \omega_j + \mathrm{Im} \left[ H \mathrm{e}^{-i\theta_j} \right] \ ,
\end{equation}
where the phase dynamics of the $j$-th oscillator ($j=1,\dots,N$) depends on its natural frequency $\omega_j$ and a driving complex-valued field $H$.
The latter can depend on time $t$, on the mean field $z(t) = \sum_{j=1}^N \mathrm{e}^{i\theta_j(t)}$, and on other auxiliary variables, but not on the (index of) oscillator, i.e. it remains identical for all oscillators $j = 1,\dots, N$.
Given the right-hand side of \eqref{single1}, the oscillators are \textit{sinusoidally coupled}.

In the thermodynamic limit ($N\to\infty$) the OA ansatz yields solutions for the dynamical evolution of the corresponding distribution function (of all the oscillators), which are attracted towards a reduced manifold of states \cite{OttAntonsen2008, OttAntonsen2009}.
Central to this is the description of the system via its distribution density $\rho(\theta, \omega, t)$.
The quantity $\rho(\theta, \omega, t)\,d\theta\,d\omega$ is the fraction of oscillators whose phases are in the range $[\theta, \theta\!+\!d\theta]$ and  have natural frequencies in $[\omega, \omega\!+\!d\omega]$ at time~$t$.
The distribution function $\rho$ obeys the continuity equation
\begin{equation}
\label{conteq}
\partial_t \rho + \partial_\theta \left(  \rho \nu \right) = 0
\end{equation}
with velocity field
\setcounter{eq_dummy}{1}%
\renewcommand{\theequation}{\arabic{equation}\alph{eq_dummy}}%
\begin{equation}
\label{velofield}
\nu(\theta, t) = \omega	+ \mathrm{Im} \left[ H(t) \mathrm{e}^{-i\theta} \right] \ .
\end{equation}
The latter can equivalently be written as\cite{MarvelMirolloStrogatz2009,Wagemakers2014}
\addtocounter{equation}{-1}\addtocounter{eq_dummy}{1}
\begin{equation}
\label{velofield2}
\nu(\theta, t )  = f\mathrm{e}^{i\theta} + h + f^\ast \mathrm{e}^{-i\theta} \ .
\end{equation}
\renewcommand{\theequation}{\arabic{equation}}%
In agreement with the assumptions on $H$ we require that the functions $f$ and $h$ may explicitly depend on time $t$, on the (now continuum form of the) mean field $z(t) = \int_{-\infty}^\infty \int_0^{2\pi} \rho \mathrm{e}^{i\theta} d\theta d\omega$, and on other auxiliary variables, but not on the the phase $\theta$ itself.

Asymptotic attractiveness of the OA manifold, given by distribution functions of the form
\begin{equation}
\label{OAmanifold}
\rho(\theta,\omega, t) = \frac{g(\omega)}{2\pi} \left\lbrace 1 + \left[ \sum_{n = 1}^\infty \alpha(\omega,t)^n \mathrm{e}^{in\theta} + \mathrm{c.c.} \right] \right\rbrace \ 
\end{equation}
that satisfy the normalization condition
\begin{equation}
\label{normalization}
\int_{-\infty}^\infty \int_{0}^{2\pi} \rho(\theta, \omega, t) \ d\theta \, d\omega = 1 \ ,
\end{equation}
has been proven for continuous frequency distribution functions $g(\omega)$ of non-zero width and for $H$ being independent of $\theta$; $\mathrm{c.c.}$ stands for complex conjugate.
Other requirements include $|\alpha(\omega,t)| \leq 1$, and some analytic continuity conditions.\cite{OttAntonsen2008, OttAntonsen2009}

In what follows we extend this approach by rigorously proving the asymptotic attractiveness of the OA manifold in the case of $H$ and $\omega$ depending on an additional parameter $\eta$ that may also influence $\theta$.
Equivalently, we include a time- and $\eta$-dependence of $f$ and $h$ in \eqref{velofield2}.
By this, we allow for an intrinsic relation between $\theta, H$, and $ \omega$, or $\theta, f$, and $h$, respectively.
As of today, the attractiveness of the OA manifold in the (time- and) parameter-dependent case has only been hypothesized\cite{MarvelMirolloStrogatz2009,PikovskyRosenblum2011} but not proven.

\section{Parameter-dependent systems}
\label{PDS}
When including additional parameters at the oscillator level, the dynamics \eqref{single1} becomes
\begin{equation}
\label{single1new}
\dot{\theta_j} = \Omega(\omega_j,\eta_j)  + \mathrm{Im} \left[ H(\eta_j,t) \  \mathrm{e}^{-i\theta_j} \right] \ .
\end{equation}
The natural frequency $\Omega$ of oscillator $j$ may therefore deviate from $\omega_j$, which promotes further heterogeneity among oscillators. 
Moreover the driving field $H$ may depend on $\eta_j$.
The right-hand side of \eqref{single1new} expresses a certain dependence on the (index of the) $j$-th oscillator.
Hence, such a dependence is no longer exclusive to the sinusoidal coupling, but also affects the natural frequency $\Omega(\omega_j,\eta_j)$ and the driving field $H(\eta_j,t)$ .

When considering $\eta$ a random variable, we may regard $\eta_j$ to be drawn from a distribution function $g(\eta)$.
Likewise $\omega_j$ may be drawn from a (different) distribution function.
The oscillator-specific parameter $\eta_j$ may change this distribution function in the oscillator's favor.
Therefore, we here incorporate a joint distribution $g(\omega, \eta)$ in the normalization condition \eqref{normalization}.
In general, $\omega$ and $\eta$ are not independent and the joint distribution consists of two nested distributions. We hence replace $\Omega(\omega_j,\eta_j)$ by $\omega(\eta_j)$. Then, in the continuum limit \eqref{single1new} reads:
\begin{equation}
\label{velofieldnew}
\partial_t \theta(\eta, t) = \omega ( \eta, t) + \mathrm{Im} \left[ H (\eta, t) \ \mathrm{e}^{-i \theta} \right] \ .
\end{equation}
The relation through $\eta$ becomes now even more evident as the temporal derivative of $\theta$ has become partial.

Again, one can introduce a distribution function $\rho(\theta, \omega, \eta, t)$, which now additionally depends on $\eta$.
And again, this distribution function satisfies the continuity equation \eqref{conteq} with velocity field \eqref{velofieldnew}.
In line with the parameter-independent case, in which the distribution function $g(\omega)$ of the natural frequencies $\omega$ had non-zero width \cite{OttAntonsen2008, OttAntonsen2009}, we assume that the distribution function $g(\eta)$ of the parameter $\eta$ also has non-zero width.
The frequency $\omega$, thus, cannot be constant but depends on $\eta$.
Likewise, the driving field $H$ depends on $\eta$.
Importantly, these two terms exhibit so an implicit dependence on $\theta$, such that the proof for the attractiveness of the OA manifold as has been derived in Ref.~\onlinecite{OttAntonsen2009} may no longer hold.
However, there is strong numerical incentive that the OA manifold fully covers the long-term behavior of the dynamics of the population of parameter-dependent phase oscillators; see, e.g., Refs.~\onlinecite{MontbrioPazoRoxin2015, PikovskyRosenblum2011, luke2013complete, so2014networks, laing2014derivation,Byrne2016,Coombes2016,Laing2016, MontbrioPazoShear2011, PazoMontbrioShear2011, MontbrioPazoColl2011, Iatsenko2013, Iatsenko2014}.

In the following we demonstrate the proof of this conjecture for a particular class of parameter-dependent systems.
We consider $\eta$ to follow a Lorentzian distribution and assume that $\omega$ depends linearly on $\eta$, i.e. $\omega(\eta,t) = a \cdot \eta + c$, where, without loss of generality, we set $a = 1$ and consider $c = c(t) \in L_1(\mathbb{R})$ an integrable, and in particular piecewise smooth, function.
Our line of argument follows closely that of Ott and Antonsen \cite{OttAntonsen2009} but we extend their results whenever necessary.
We would like to note that our findings remain valid for a larger class of distribution functions as has been depicted in detail in Ref.~\onlinecite{OttAntonsenComment}.
We will comment on this and consider more general $\eta$-dependencies of $\omega$ in Sections \ref{general} and \ref{general2}.

Let $g(\eta)$ be a Lorentzian centered around $\eta = \eta_0$ with width $\Delta$, i.e. $g(\eta)\!\sim\!L(\eta_0, \Delta)$.
For the aforementioned linear dependency $\omega(\eta,t) = a \cdot \eta + c$, we have $\tilde g(\omega) = \hat g(\eta)\!\sim\!L(\eta_0+c,\Delta)$
with frequency $\omega = \omega(\eta)$ that, in general, will depend on $\eta$.
In this case $\omega$ is fully described by (the distribution of) $\eta$ and  
the distribution density reduces to $\rho(\theta,\omega,\eta, t) = \rho(\theta,\eta,t)$.\cite{Note2}
%\footnote{Alternatively, the dependence $\omega(\eta)$ may be constituted by considering $\rho$ as a conditional probability density $\rho(\theta, \omega, t | \eta) = \rho(\theta, \omega | \eta, t)$ in line with Ref.~\onlinecite{MontbrioPazoRoxin2015}.} 
This can be expanded as a Fourier series in $\theta$ similar to Eqs.(5~\&~6) in Ref.~\onlinecite{OttAntonsen2009},
where it is further decomposed into $\rho(\theta,\eta, t) = \hat g(\eta) / (2\pi) \cdot [1 + \rho_+(\theta, \eta, t) + \rho_-(\theta, \eta, t)]$.
Next to the assumption that the analytic continuation of $\rho_+$ ($\rho_-$) into $\mathrm{Im} (\theta)\!>\!0$ ($\mathrm{Im} (\theta)\!<\!0$) has no singularities and decays to zero as $\mathrm{Im} (\theta) \to +\infty$ ($\mathrm{Im} (\theta) \to -\infty$), we exploit the symmetry of the Fourier expansion and focus on $\rho_+$.
In particular, we expect $\rho_+$ to fulfill these conditions initially, i.e. $\rho_+(\theta, \eta, 0)$ can be continued into the complex $\eta$-plane, is analytic in $\mathrm{Im}(\eta) < 0 $ and decays to zero for $\mathrm{Im}(\eta) \to -\infty$. These conditions are satisfied for all $t>0$.\cite{OttAntonsen2008}

We can further decompose $\rho_+$ into two parts, $\rho_+ = \hat \rho_+ + \hat \rho'_+$, where $\hat \rho'_+$ lies on the OA manifold and follows the dynamics given by Eq.(9) in Ref.~\onlinecite{OttAntonsen2009}.
For the sake of completeness, this dynamics prescribes the evolution of the Fourier coefficients $\hat \rho'_+$ to the form $\hat \rho'_n(\eta,t) = \left[ \alpha(\eta,t)\right]^n$, and reads
\begin{equation}
\partial_t \alpha + i \eta \alpha  + \frac{1}{2} \left( H\alpha^2 - H^\ast \right) = 0 \ . 
\end{equation}
The quantity $\hat \rho_+$, on the other hand, is a solution of
\begin{equation}
\label{conteqrho+}
\partial_t \hat \rho_+ + \partial_\theta \left\lbrace \left[ \omega + \frac{1}{2i} \left( H \mathrm{e}^{-i\theta} - H^\ast \mathrm{e}^{i\theta} \right) \right] \hat \rho_+ \right\rbrace = 0 \ .
\end{equation}
Both the frequency $\omega$ and the field $H$ may depend explicitly on $\eta$.
To guarantee that the dynamics \eqref{single1new}, whose state at time $t$ can be represented by the afore-defined order parameter $z(t)$ in its continuous form,
\begin{equation}
\label{orderparam}
z(t) = \int_{-\infty}^\infty \int_0^{2\pi} \rho(\theta, \eta, t) \mathrm{e}^{i\theta} d\theta d\eta \ ,
\end{equation}
is asymptotically attracted by the OA manifold, it suffices to show that
\begin{equation}
\label{Claim}
\lim_{t \to +\infty} \int_{-\infty}^{+\infty} \hat \rho_+ (\theta, \eta, t ) \hat g(\eta) d\eta =  0 
\end{equation}
holds. Before showing this, however, we would first like to remark that, without loss of generality, the center of the Lorentzian frequency distribution $\hat g(\eta) \sim L(\eta_0 + c, \Delta)$ can be considered zero since we may introduce a change of variables, $\tilde \theta = \theta - (\eta_0 t + C(t) )$,
where $C(t)$ is an antiderivative of $c(t)$.
Furthermore, we can adjust \eqref{Claim} by substituting $\hat g$ by $g$.

If $\hat \rho_+$ is analytic in the lower half $\eta$-plane and decays to zero as $\mathrm{Im} (\eta) \to -\infty$, one can multiply \eqref{conteqrho+} by $g(\eta)d\eta$ and integrate the result by employing the residue theorem.
Hence, the integrals can be evaluated at the residue of the enclosed pole of $g(\eta)$ at $\eta = - i \Delta$. We find
\begin{align*}
&\partial_t \hat \rho_+(\theta, -i\Delta, t) + \partial_\theta \Biggl\{ -i\Delta \cdot \hat \rho_+(\theta, -i\Delta, t) \\
&\hspace*{2em}+ \frac{1}{2i} \left[ \int_{-\infty}^{+\infty} H(\eta,t)\hat \rho_+(\theta, \eta, t) g(\eta) d\eta \ \mathrm{e}^{-i\theta} 
%\right. \\
 %&\hspace*{2em}\left. 
  - \int_{-\infty}^{+\infty} H^\ast(\eta,t)\hat \rho_+(\theta, \eta, t) g(\eta) d\eta \ \mathrm{e}^{i\theta} \right]\Biggr\} = 0 \ .
\end{align*}
The two remaining integrals can be determined provided that $H$ and $H^\ast$ have no singularities in the lower half $\eta$-plane and do not increase ``too'' fast for $\mathrm{Im} (\eta) \to -\infty$. 
Since $g$ is a Schwartz function, we only need $H$ to diverge at most sub-exponentially.
For common choices of $H$, as listed in Ref.~\onlinecite{OttAntonsen2009}, these requirements are met indeed, which yields
\begin{align}
\label{eq17}
&\partial_t f_+(\theta, t) + \partial_\theta \left[ v(\theta, t) f_+(\theta, t) \right]  = 0 \ ,\\
\label{eq18}
&v(\theta, t)  =  - i \left[ \Delta + \frac{1}{2} \left( \mathrm{e}^{-i\theta}  H(t) - \mathrm{e}^{i\theta} H^\ast(t) \right) \right] \ .
\end{align}
Here we substituted $f_+(\theta, t) = \hat \rho_+(\theta, -i\Delta, t)$ and $H(t) = H(-i\Delta, t)$.
These equations agree exactly with Eqs.(17~\&~18) in Ref.~\onlinecite{OttAntonsen2009}.
Hence, following the same reasoning around Eqs.(19-31) in Ref.~\onlinecite{OttAntonsen2009} one can conclude that \eqref{Claim} is fulfilled.
To underscore the line of argument, we would like to give a short sketch of the proof.
First, by introducing a conformal transformation of the upper half complex $\theta$-plane into the unit disc via $w = \mathrm{e}^{i\theta}$, one can rewrite (\ref{eq17}~\&~\ref{eq18}) as
\begin{equation}
\label{chareq}
\frac{d}{dt}\tilde f_+(w, t) + \tilde f_+(w, t) \partial_w \tilde v(w,t) = 0 \ ,
\end{equation}
where $\tilde f_+$ and $\tilde v$ are the transformed functions from (\ref{eq17}~\&~\ref{eq18}), and $d/dt = \partial/\partial t + \tilde v \partial/\partial w$.
\eqref{chareq} can be integrated using the method of characteristics for linear and homogeneous partial differential equations\cite{Evans1998}.
Here we require $\tilde f_+ \in C^2(\mathbb{R})$ but $\tilde v$ does not need to be continuous.
This yields
\begin{equation}
\label{charsol}
\tilde f_+(w, t) = \tilde f_+(W(w,0), 0) \exp \left[ -\mu (w,t) \right] \ , 
\end{equation}
as solution with 
\begin{equation}
\label{charhelp}
\mu(w,t) = \int_0^t \left. \partial_{w'} \tilde v(w',t') \right|_{w' = W(w,t')} dt' \ ,
\end{equation}
and the characteristics are given by 
\begin{equation}
\label{charorbits}
\partial_{t'}W(w,t') = \tilde v(W(w,t'),t') \ ,
\end{equation}
with final condition $W(w,t) = w$.
Finally, in order to show that $\tilde f_+(w,t) \to 0$ for $t \to \infty$, which, by \eqref{charsol}, we prove that
\begin{equation}
\label{charlim}
\lim_{t \to \infty} \mathrm{Re} \ \left[ \mu(w,t) \right] = + \infty \ .
\end{equation} 
The details for the rather lengthy computation can be found in Ref.~\onlinecite{OttAntonsen2009}.
We here we would only like to mention that the integral in \eqref{charhelp} is split into three distinct parts, each of which is evaluated and while two of them remain bounded, the third diverges at the rate $\Delta t$, presuming $\Delta > 0$.
This eventually completes the proof and underlines the importance that the distribution function $g(\eta)$ must have non-zero width $\Delta$.
We would also like to note that in the final step of the proof the continuity of $v$ is required, i.e. $H$ in \eqref{eq18} must be continuous.
If one includes, e.g., square functions in the time-dependent parts of the frequency term and/or driving field, one is confronted with jump discontinuities, which become present in the right-hand side of \eqref{eq18} either directly or indirectly via the order parameter $z(t)$.
A closer look at Ref.~\onlinecite{OttAntonsen2009}, however, confirms that for small jumps the reasoning can be guaranteed and for proper choices of a time constant $T$ their Eq.$(31)$ holds.
Thus, we can argue that OA attractiveness will be maintained even in the case of discontinuities, which also confirms our rather long assumption for $c(t)$ to be in $L_1(\mathbb{R})$ in the linear dependence of $\omega(\eta) = a\eta + c$.

So far we only considered a Lorentzian distribution and some linear dependence of $\omega$ on $\eta$.
However,
our result can be  extended to a much broader class of distribution functions $g(\eta)$, non-linear dependencies $\omega(\eta)$, or even joint distributions $g(\omega, \eta)$ in the case of $\Omega(\omega, \eta)$; see Section \ref{general} below.
Hence, it is proper to say that the asymptotic attractiveness of the OA manifold for parameter-dependent systems of coupled phase oscillators is generic.
Note that the proof remains identical if $\theta = \theta(t)$ does not depend on the parameter $\eta$, that is, when there is no correlation between specific oscillators and their dynamics.
We call this case ``weak'' parameter-dependence, which has been coined in several earlier studies, e.g., Refs.~\onlinecite{MarvelMirolloStrogatz2009,So2011,Wagemakers2014,PikovskyRosenblum2011}, where parameters were introduced as auxiliary variables.
Our result therefore confirms the attractiveness of the OA manifold also in this case, as has simplifyingly been taken for granted in the afore-cited studies.

\section{Networks of QIF and theta neurons}
\label{QIFsection}
As mentioned above, there is a variety of recent papers that showed numerically how the dynamics of networks of theta neurons is time asymptotically attracted by the OA manifold \cite{luke2013complete, so2014networks, laing2014derivation}.
Recently, Montbri\'o and co-workers studied how the macroscopic dynamics of a network of quadratic integrate-and-fire (QIF) neurons is  described by a low-dimensional system by using a so-called Lorentzian ansatz \cite{MontbrioPazoRoxin2015}. 
By transforming the QIF neurons into a network of theta neurons, their Lorentzian ansatz does resemble the OA ansatz with parameter-dependent frequency and driving field, as considered in Section \ref{PDS}.

To be more precise, the dynamics of the membrane potential $V_j$ of a QIF neuron may be described by
\begin{equation}
\label{QIF}
\dot V_j = V_j^2 + I_j \ , \quad \text{if } V_j \geq V_p \ , \text{ then } V_j \leftarrow V_r \ ,
\end{equation}
for $j = 1, \dots ,N$.
Here, $I_j$ denotes an input current, $V_p$ a peak value, and $V_r$ a reset value.
Once the membrane potential $V_j$ reaches $V_p$, the neuron emits a spike, and $V_j$ will be reset to $V_r$.
Commonly, the limit $V_p = - V_r \to \infty$ is considered.
The input current $I_j$ consists of a neuron-specific quenched component $\eta_j$, a common time-dependent input $I(t)$ and a coupling term $J s(t)$, combining the synaptic weight $J$ and a smooth mean synaptic activation $s(t)$, resulting in
\begin{equation}
\label{inputcurrent}
I_j = \eta_j + J s(t) + I(t) \ .
\end{equation}
The latter two time-dependent components are identical for all neurons in the network.
In order to describe the macroscopic behavior of the network, Montbri\'o and co-workers
used the Lorentzian ansatz
\begin{equation}
\label{Lorentzian}
\rho(V | \eta, t ) = \frac{1}{\pi} \frac{x(\eta,t)}{\left[ V - y(\eta,t)\right]^2 - x(\eta,t)^2} \ ,
\end{equation}
with center $y(\eta,t)$ and time-dependent half-width $x(\eta, t)$,
which turns out to exhibit the long-term solution for the distribution of the membrane potentials.
The properties $x(\eta,t)$ and $y(\eta,t)$ that define the distribution function \eqref{Lorentzian} are also closely linked to the firing rate of the neuronal population and to the mean membrane potential, respectively.
While the Lorentzian ansatz applies to the (membrane voltage) dynamics of QIF neurons, we are here primarily interested in the phase dynamics.
Using $V_j = \tan(\theta_j/2)$ one can transform (\ref{QIF}~\&~\ref{inputcurrent}) into theta neurons,\cite{ermentrout1986parabolic}
\begin{equation}
\label{thetaneuron}
\dot \theta_j = ( 1\!-\!\cos \theta_j ) + ( 1\!+\!\cos \theta_j ) \left[ \eta_j + J\!\cdot\!s(t) + I(t) \right] \ .
\end{equation}
In \eqref{thetaneuron} the time-independent injected current $\eta_j$ is drawn from a distribution function $g(\eta)$.
For the sake of legibility we abbreviate the non-autonomous part of \eqref{thetaneuron} as
\[
J\!\cdot\!s(t) + I(t) = c(t) - 1 \ .
\]
Rearranging terms and considering the thermodynamic limit, one can rewrite \eqref{thetaneuron} as
\begin{equation}
\label{thetaOA}
\partial_t \theta(\eta, t) = \nu(\theta, \eta, t) =  \Omega ( \eta, t) + \mathrm{Im} \left[ H (\eta, t) \mathrm{e}^{-i \theta} \right] 
\end{equation}
with $H (\eta, t) = i(-1+\eta+Js+I)= i(\eta+c-2)$ and $\Omega ( \eta, t) =  \eta+c$; cf. Ref.~\onlinecite{MontbrioPazoRoxin2015}.

To apply our result from above, one has to show that $H$ does not diverge exponentially when $\mathrm{Im} (\eta) \to -\infty$, and that $c(t)$ possesses an antiderivative.
On the one hand, for the components of $c(t)$ with $s(t)$ being smooth and $I(t)$ piecewise smooth and integrable, there will always exist an antiderivative of $c(t)$.
On the other hand, we have $H(\eta) =  i \eta + const$, such that $H$ grows only linearly for $\mathrm{Im} (\eta) \to -\infty$.
Hence, we find that the OA manifold does asymptotically attract the macroscopic behavior of a network of coupled theta neurons. 
Due to the existence of a conformal mapping between the quantity $w(\eta,t) = x(\eta,t) + i y(\eta,t)$ and the function $\alpha(\eta,t)$ defining the OA manifold \eqref{OAmanifold}\cite{Note3}
%\footnote{We substituted $\omega $ by $\eta $ in line with our arguments in Section \ref {PDS}. However, we do assume an implicit dependence $\omega = \omega (\eta )$.}
, see also Eq.(15) in Ref.~\onlinecite{MontbrioPazoRoxin2015}, we have also proven the attractiveness of the Lorentzian ansatz \eqref{Lorentzian} for a network of QIF neurons.

\section{General parameter distributions}
\label{general}

As already mentioned in Section \ref{PDS}, the assumptions of a linear relation between $\omega$ and $\eta$ and of $\eta$ being drawn from a Lorentzian can be loosened in many respects. We first consider $g(\eta)$ to still be a Lorentzian centered around $\eta = \eta_0$ with width $\Delta$, i.e. $g(\eta)\!\sim\!L(\eta_0, \Delta)$.
The linear dependency $\omega(\eta,t) = a \cdot \eta + c$ may be generalized by considering both $a=a(t)$ and $c=c(t)$ time-dependent. 
Then, by the common transformation properties for Lorentzian (Cauchy) distributions, $\omega$ follows a Lorentzian of the form $g(\omega)\!\sim\!L(a \eta_0 + c, \Delta |a|)$.
Let $a \neq 0$ be constant.
Then a similar change of variables, $\tilde \theta = \theta - (a\eta_0 t + C(t))$,
with $C(t)$ being the antiderivative of $c(t)$, keeps the distribution function centered around $0$.
Without loss of generality we set $a=1$;
even if $a = a(t)$ and $a(t)>0$ or $a(t) < 0 $ for all $t>0$, the rescaling of $\theta$ retrieves that we can stick to our assumption $a = 1$.
If, however, $a$ changes sign at, e.g., $t = t_0$, then the scale parameter $\Delta |a|$ tends to zero for $t \to t_0$.
Due to \eqref{OAmanifold} also $\rho(\theta,\omega,t)$ will exhibit a $\delta$-peak at $t = t_0$. 
In this case our results are not readily applicable\cite{OttAntonsenComment}.
However, if $\dot a(t_0) \neq 0$, then we can shift the initial time to zero, $t_0 \mapsto 0$.
Whenever $\rho_+(\theta,\omega,t_0)$ satisfies the necessary initial conditions, the OA manifold will remain attracting for all $t > t_0$, given that $t_0 = \max \{t\in \mathbb{R} \ | \ a(t) = 0 \}$.

We proceed with more general cases of frequency and parameter distributions.
In Ref.~\onlinecite{OttAntonsenComment}, the authors elegantly extend the original proof, which considers only Lorentzian frequency distributions:
Instead of demanding analytic continuity of both the frequency distribution $g(\omega)$ and the initial condition into the whole lower $\omega$-half plane, it suffices that $g$ and the initial condition have analytic continuations into a strip $S$ defined by $0 \geq \mathrm{Im} (\omega) > - \sigma$ and $ -\infty \leq \mathrm{Im} (\omega) \leq +\infty$ with $\sigma > 0$, where neither of them has singularities and both approach zero as $|\omega| \to \infty$.
Thereby the class of applicable distribution functions includes Gaussians, sech-distributions, and many more, and even multimodal distributions can be incorporated as long as these functions have finite non-zero widths; see references in Ref.~\onlinecite{OttAntonsenComment}.
This approach can be adopted and used in our $\eta$-parameter-dependent case.
For this let us assume again individual oscillators given by \eqref{single1new}.
As mentioned in Section \ref{PDS}, we might be confronted with a nesting of the distributions $\tilde g(\omega)$ and $g(\eta)$ for $\omega$ and $\eta$.
In particular, the latter may determine the first in an oscillator-specific way.
That is the reason why the resulting distribution function $\hat g(\eta)$ can become arbitrarily complicated.
However, as long as the analytic continuations of $\tilde g$ and $g$ into the strip $S$ (for some $\sigma > 0$ as defined above) do not have singularities, and neither $\tilde g$ nor $g$ features a $\delta$-peak in their time-evolutions, 
also $\hat g$ will behave as required.
An additional requirement is that the product $H(\eta, t) \hat g(\eta)$ satisfies these conditions, too.
This means that we have to find a strip $S' \subset S$, defined by $0 < \sigma' \leq \sigma$, in which $H \hat g$ has an analytic continuation, does not have singularities, its time evolution does not feature $\delta$-peaks (if necessary we have to reset the initial time point after such a peak), and that we require $|H(\eta_r + i \eta_i,t) \hat g(\eta_r + i \eta_i)| \to 0$ for $|\eta_r| \to \infty$ and $0 > \eta_i > -\sigma'$.
In particular, $H$ must not grow faster than $\hat g$ decays, such that the OA manifold continues to capture the long-term dynamics of the system.

Revisiting the example from Section \ref{QIFsection}, where $H(\eta) = i(\eta + c - 2)$ and $\hat g(\eta) \sim \mathrm{L}(\eta_0 + c, \Delta)$, we find that $\hat g$ decays exponentially for $|\eta_r| \to \infty$ such that $H$ must not increase at an exponential rate.
In fact, $H$ dot not have any singularities in the whole complex $\eta$-plane (except for $|\eta| \to \infty$), and $H(\eta_r + i \eta_i) = - \eta_i + i \eta_r + const = \mathcal{O} (\eta_r)$ for $|\eta_r| \to \infty$.
Consequently, for large $|\eta_r|$, the product $H \hat g$ will be dominated by $\hat g$ such that all assumptions are fulfilled.
Hence, we can confirm again the attractiveness of the OA manifold.\cite{Note4}
%\footnote{To give a brief idea of the proof, it is important to note that, next to the assumption that in $S'$ the product $H \hat g$ decays to zero for $|w|\to \infty$, the crucial point for proving the attractiveness of the OA manifold is that $\sigma' > 0 $. To be more precise, given the integral expression (equivalent to) \eqref{Claim}, the idea is to shift the path of integration from the real $\eta$-axis to the line $\eta_r + i \eta_i$ with $0 > \eta_i > -\sigma'$, $-\infty \leq \eta_r \leq \infty$, for details see Ref.~\onlinecite{OttAntonsenComment}. This leads directly to (\ref{eq17}~\&~\ref{eq18}) from where one can complete the proof along the known formalism outlined in Section \ref{PDS}.}

We would like to remark that initial conditions on the oscillator distribution function, $\rho(\theta, \eta, 0)$, play an important role.
If they fail to be satisfied, this may hinder the OA manifold to attract the dynamics.
For an example we would like to refer to Appendix C of Ref.~\onlinecite{OttPlatigAntonsenGirvan2008}, in which the specific time point has to be determined appropriately in order to set up promising initial conditions.

\section{Applications -- realistic settings}
\label{general2}
So far, we only considered non-independent frequency and parameter distributions, $\tilde g(\omega)$ and $g(\eta)$, respectively.
In general, however, one cannot take this ``simple'' dependence for granted.
The additional parameter might be multi-dimensional, i.e. $\eta \in \mathbb{R}^n$ with $n > 1$.
When considering the thermodynamic limit of infinitely many coupled oscillators, the dynamics \eqref{single1new} may obey
\begin{equation}
\label{single1newthermo}
\partial_t \theta(\eta, t) = \Omega (\omega, \eta, t) + \mathrm{Im} \left[ H (\eta, t) \ \mathrm{e}^{-i \theta} \right] \ .
\end{equation}
Employing the OA ansatz for this system one has to encounter distribution functions given like
\begin{equation}
\begin{gathered}
\rho(\theta,\omega,\eta, t) = \frac{g(\omega,\eta)}{2\pi} \left\lbrace 1 + \left[ \sum_{k = 1}^\infty \alpha(\omega,\eta,t)^k \mathrm{e}^{ik\theta} + \mathrm{c.c.} \right] \right\rbrace \\ 
\int_{\mathbb{R}^n} \int_{-\infty}^\infty \int_{0}^{2\pi} \rho(\theta, \omega, t) \ d\theta \ d\omega \ d\eta = 1 \ ;
\end{gathered}
\label{OAmanifoldfurgen}
\end{equation}
the joint distribution $g(\omega,\eta)$ is a major modification to the setting considered before.
Does the OA manifold remain attracting?
\eqref{single1newthermo} suggests the phase $\theta = \theta(\eta, t)$ to depend on the parameter $\eta$ in line with our notion of parameter-dependent systems.
But it is unclear whether the OA manifold is attracting even without this particular correlation between phase, natural frequency, and driving field.
If, however, the OA attractiveness can be proven for systems with generalized natural frequency $\Omega$ and driving field $H$ as in \eqref{single1newthermo}, this will allow for a further and even broader extension of the existing theory.
In the following we first list a few examples for which numerical simulations have been reported and that give strong incentive that the OA ansatz may indeed be valid.
We will show how our proof can be adopted, thereby confirm the OA attractiveness, and set the numerical results on solid ground.
Last, we provide some general properties of $\Omega$ and $H$ for which the OA ansatz holds. 

We start with the Winfree model\cite{Winfree1967} which is an early mathematical description of synchronization phenomena in large populations of biological oscillators.
Rewritten in terms of \eqref{single1newthermo} this model takes the form 
\begin{equation}
\begin{gathered}
\partial_t\theta  = \Omega(\omega, \eta, t) + \mathrm{Im} \left[ H(\eta, t) \mathrm{e}^{-i\theta} \right] \\
\Omega(\omega, \eta, t)  = \omega + \sigma \eta(t),  \ \
H(\eta, t)  = \mathrm{e}^{-i\beta} \eta(t), \ \ \text{and}\ \
\eta(t)  = \varepsilon h(t) \ ,
\end{gathered}
\label{Winfree_intro}
\end{equation}
where $h(t)$ is a smooth function depending only on the mean field $z(t)$ but not on the phase itself\cite{PazoMontbrio2014}.
In particular, this model contains time-dependent parameters, see also Ref.~\onlinecite{PetkoskiPRE2012}.

Next, we consider reaction-diffusion systems with heterogeneous, self-oscillating elements.
In particular, we study the mean-field version of the complex Ginzburg-Landau equation, whose equation describes a population of globally coupled limit-cycle oscillators.
Hence, we can rewrite the dynamics in form of \eqref{single1newthermo}.
By introducing a shear (or nonisochronicity) parameter $\eta $ as an additional random variable and transforming the system through a phase reduction, the governing equations in the continuum limit read\cite{MontbrioPazoShear2011,PazoMontbrioShear2011,MontbrioPazoColl2011}:
\begin{equation}
\begin{gathered}
\partial_t \theta  = \Omega(\omega, \eta, t) + \mathrm{Im} \left[ H(\eta, t) \mathrm{e}^{-i\theta} \right] \\
\Omega(\omega, \eta, t) = \omega + K \eta \ \ \text{and} \ \ 
H(\eta, t)  = K  z (1 - i \eta) \ ,
\end{gathered}
\end{equation}
where $K$ denotes the coupling strength and $z=z(t)$ is the order parameter. The frequency $\omega$ and the shear $\eta$ are drawn from a joint distribution $g(\omega,\eta)$.
In contrast to Section \ref{PDS}, we explicitly allow the additional parameter $\eta$ to be drawn from another frequency distribution.
For the joint distribution one has to address two scenarios.
Either, the random variables are independent, such that the joint distribution can be split into $g(\omega, \eta) = g_1(\omega) g_2(\eta)$, or they are not independent.
Iatsenko and co-workers, who independently investigated the Kuramoto model with both distributed natural frequencies and distributed coupling strengths, i.e. with two random variables $\omega$ and $\eta$, coined the first case as \textit{uncorrelated} joint distributions, and the latter as \textit{correlated}, see Ref.~\onlinecite{Iatsenko2013,PetkoskiPRE2013,Iatsenko2014}.
Furthermore, frequency-weighted coupling\cite{WangLi2011,Xu2016}, i.e. the driving field additionally depends on $\omega$, $H = H(\omega, \eta, t)$, can be approached with the formalism introduced above.

Last but not least, the upcoming branch of heterogeneous mean fields\cite{Vespignani2012} falls in a category whose mean field dynamics can be described along the OA ansatz.
The heterogeneous mean field approach deals with networks that are not all-to-all coupled but they exhibit some particular (and sparse) network topology, and therefore can barely be studied analytically.
Given a network with a particular degree distribution, however, it is possible to introduce so-called degree-block variables, whose dynamics govern the evolution of all nodes which have the same degree $k$.
This approach reveals the same equations as the annealed networks approximation\cite{Rodrigues2016,Doro2008}, which can hence be considered equivalent.
Recent studies considered the heterogeneous mean fields of the Kuramoto model, e.g., on scale-free\cite{Coutinho2013,Yoon2015,Lopes2016} and random Erd\"{o}s-R\'{e}nyi networks\cite{Coutinho2013}.
The starting point is a specifically coupled Kuramoto network with coupling strength $K$ and adjacency matrix $A = (a_{ij})$ with $i,j=1,\dots, N$,
\begin{equation}
\dot \theta_j = \omega_j + K \sum_{k=1}^N a_{jk} \sin(\theta_k - \theta_j) \ .
\end{equation}
We can cluster various node dynamics by replacing the adjacency term with an expectation value for their node degree $\eta_j$.
Ideally, the underlying topology exhibits some well-defined degree distribution $P(\eta)$.
In the continuum limit $N \to \infty$, these node degrees are substituted in the phase dynamics as weighted, distributed coupling strengths, so that the governing dynamics read
\begin{equation}
\begin{gathered}
\partial_t \theta(\eta,t)  =  \ \Omega(\omega, \eta, t) + \mathrm{Im} \left[ H(\eta, t) \mathrm{e}^{-i\theta} \right] \\
\Omega(\omega, \eta, t)  = \omega \ \ \text{and} \ \
 H(\eta, t)  = K \eta z(t) \ ,
\end{gathered}
\end{equation}
where $\omega$ and $\eta$ are drawn from a joint distribution $g(\omega, \eta) = P(\eta) g_1(\omega)$.
This setup is amenable to, e.g., random fields, as has been presented in Ref.~\onlinecite{Lopes2016} where oscillators are enforced through local fields, which find their way into the specific forms for $\Omega$ and $H$.

In all these different classes of parameter-dependent networks, we will show how the OA attractiveness can be regained.

\subsection{Winfree model}
As said, the Winfree model describes macroscopic synchronization phenomena of large oscillator systems whose individual nodes are naturally pulse-coupled with one another.
The introduction of phase response curves (PRC) allows for quantifying how the phase of an oscillator responds to the pulse-like perturbations from the other oscillators.
The general form of model reads at the single node level
\begin{equation}
\label{Winfree_model}
\dot \theta_j = \omega_j + Q(\theta_j) \frac{\varepsilon}{N} \sum_{k=1}^N P(\theta_k) \ ,
\end{equation} 
where $\varepsilon$ denotes the coupling strength, $Q$ is the PRC and $P$ is a pulse-like signal.
Following the notation of Paz\'{o} and Montrbri\'{o} in Ref.~\onlinecite{PazoMontbrio2014}, we consider PRCs with sinusoidal shape,
\begin{equation}
\label{Winfree_PRC}
Q(\theta) = \sigma - \sin(\theta + \beta) \ ,
\end{equation}
with an offset parameter $\sigma$, and a phase-lag $\beta$.
Moreover, we assume the pulse-like signal to be smooth,
\begin{equation}
\label{Winfree_signal}
P(\theta) =  P_n(\theta) = a_n (1 + \cos \theta)^n \ ,
\end{equation}
with $n \in \mathbb{N}_{\geq 1}$ controlling the width of the pulses, and $a_n$ is a normalizing constant.
In the thermodynamic limit, we regain \eqref{Winfree_intro} as
\begin{equation}
\label{Winfree_thermo}
\partial_t\theta = \omega + \varepsilon \sigma h(t) + \mathrm{Im} \left[ \varepsilon \mathrm{e}^{-i\beta} h(t) \mathrm{e}^{-i\theta} \right] \ ,
\end{equation}
where the coupling function incorporates the smooth mean field 
\begin{equation}
\label{Winfree_meanfield}
h(t) = h_n(t) = \int_0^{2\pi} P_n(\theta) d\theta = 1 + 2(n!)^2 \sum_{k=1}^n \frac{\mathrm{Re} (z^k)}{(n+k)!(n-k)!} 
\end{equation}
with $z$ the common (Kuramoto) order parameter \eqref{orderparam}.
The frequency $\Omega(\omega,t) = \omega + c(t)$ with $c(t) = \varepsilon \sigma h(t)$ has a form identical to Section \ref{PDS}, where $\omega$ follows a Lorentzian frequency distribution $g(\omega)$.
Since the order parameter $z(t)$ is bounded with $ | \mathrm{Re} (z) | \leq 1$, we have $h(t) \geq 0$ for all $t \geq 0$.
Furthermore, the driving field does not depend on additional parameters, so that our proof can be directly applied, confirming that the OA ansatz holds and the OA manifold indeed captures the long-term dynamics of the Winfree model.

An alternative proof for the case of time-dependent frequency and driving field can be found in Ref.~\onlinecite{PetkoskiPRE2012}.
However, as we have depicted in Section \ref{general}, our proof generalizes their findings and extends them to a broader class of frequency distribution functions $g(\omega)$.
Of particular interest in the non-autonomous extension is also the matter of discontinuities.
Recall that in Section \ref{QIFsection} we introduced a time-dependent input current $I(t)$, see \eqref{inputcurrent}, which can, e.g., take the form of a square function with jump-discontinuities.
Our proof applies to this specific feature and confirms existing numerical results\cite{MontbrioPazoRoxin2015}.

\subsection{Limit-cycle oscillations with shear}
\label{subsec_shear}

Investigating collective synchronization usually addresses networks of coupled elementary oscillatory units.
The dynamics of these units may be described as normal form 
\begin{equation}
\label{oscillator_normal}
\dot \varrho = \varrho (1 -\varrho^2) \ , \quad \dot\theta = \omega + \eta (1 - \varrho^2 )\ ,
\end{equation}
where $\varrho$ denotes the radius and $\omega$ determines the frequency of rotation on the stable limit cycle with $\varrho(t) \equiv 1$.
The parameter $\eta$ quantifies the shear, or non-isochronicity, of the flow, i.e. how strongly perturbations away from the limit cycle modify the phase dynamics.
When we consider an all-to-all coupled population of $N \gg 1$ of these oscillatory units, we arrive at the mean-field version of the complex Ginzburg-Landau equation with dissipative coupling
\begin{equation}
\label{GinzburgLandau}
\dot z_j = z_j \left[ 1 + i\left(\omega_j + \eta_j\right) - \left(1 + i\eta_j\right) \left|z_j\right|^2  \right] + \frac{K}{N} \sum_{k=1}^N\left(z_k - z_j\right) \ ;
\end{equation}
$z_j = \varrho_j \mathrm{e}^{i\phi_j}$.
Heterogeneity among the population is promoted by having the frequency $\omega_j$ and shear parameters $\eta_j$ drawn from a distribution function $g(\omega, \eta)$.
In the weakly coupled case, i.e. the coupling strength $|K|$ is small, a phase reduction allows us to describe the dynamics of the system by their phases only.
In the continuum limit $N \to \infty$, we can introduce the phase distribution function $\rho(\theta, \omega, \eta, t)$.
Note that $\omega$ and $\eta$ are independent, so that neither of them is redundant.
Accordingly, the order parameter $z$ takes now the form
\begin{equation}
\label{orderparam2}
z(t) = \int_{-\infty}^\infty \int_{-\infty}^\infty \int_0^{2\pi}  \rho(\theta, \omega, \eta, t) \mathrm{e}^{i\theta} \ d\theta d\omega d \eta \ .
\end{equation}
Thus, the phase dynamics reads 
\begin{equation}
\label{GLphases}
\partial_t\theta = \omega + K \eta + \mathrm{Im} \left[ Kz(t) (1 - i \eta) \mathrm{e}^{-i\theta} \right] \ ,
\end{equation}
and the phase distribution function satisfies the continuity equation
\begin{equation}
\label{shear_conteq}
\partial_t \rho + \partial_\theta	\left( v \rho \right) = 0 \ ,
\end{equation}
with $v$ the right-hand side of \eqref{GLphases}, see also Refs.~\onlinecite{MontbrioPazoShear2011,MontbrioPazoColl2011}.
Using the notion of \eqref{single1newthermo}, the frequency and the driving field are both time-varying and depend on the additional shear parameter $\eta$:
\begin{equation}
\label{shear_prop}
\Omega(\omega, \eta, t) = \omega + K \eta  \ , \quad H(\eta,t) = K z(t) (1 - i \eta) \ .
\end{equation}
To assure that the OA manifold indeed exhibits the mean field dynamics of this system with shear, we have to adapt our proof from Section \ref{general} for the joint distribution $g(\omega, \eta)$.

The general idea is again to decompose the distribution function $\rho$ in Fourier space into 
\begin{equation}
\rho(\theta, \omega, \eta, t) = \frac{g(\omega, \eta)}{2\pi} \left[ 1 + \rho_+(\theta, \omega, \eta, t) + \rho_-(\theta, \omega, \eta, t) \right] 
\end{equation}
and use symmetry assumptions to focus on $\rho_+$, which again will be decomposed into $\rho_+ = \hat \rho_+ + \hat \rho'_+$.
While $\hat \rho'_+$ lies on the OA manifold and has Fourier coefficients $\hat \rho'_{+,n} = \left[ \alpha(\omega, \eta, t) \right]^n$,
$\hat \rho_+$ solves
\begin{equation}
\label{shear_conteqrho+}
\partial_t \hat \rho_+ + \partial_\theta \left\lbrace \left[ \Omega(\omega, \eta, t) + \frac{1}{2i} \left( H(\eta,t) \mathrm{e}^{-i\theta} - H(\eta,t)^\ast \mathrm{e}^{i\theta} \right) \right] \hat \rho_+ \right\rbrace = 0 \ .
\end{equation}
The assumptions on the analytic continuation properties of Section \ref{general} hold -- in particular we need analytic continuations with respect to both $\omega$ and $\eta$ into strips $S_\omega$ and $S_\eta$. Hence we have to show that
\begin{equation}
\label{shear_sol}
\lim_{t\to \infty} \int_{-\infty}^\infty \int_{-\infty}^\infty \hat \rho_+(\theta, \omega, \eta, t) \ g(\omega, \eta) \ d\omega d\eta = 0 \ .
\end{equation}
Discussing general solutions of \eqref{shear_sol} given an arbitrary joint distribution function are beyond the scope of this paper.
However, for particular $g(\omega,\eta)$ we can affirm the attractiveness of the OA manifold for these parameter-dependent systems.
To begin with, we use the assumption of Montbri\'{o} and Paz\'{o} that the joint distribution can be written as the product of two Lorentzians\cite{MontbrioPazoShear2011},
\begin{equation}
\label{shear_uncorr}
g(\omega,\eta) = g_1(\omega) g_2(\eta) = \frac{\delta/\pi}{(\omega-\omega_0)^2 + \delta^2}\frac{\gamma/\pi}{(\eta-\eta_0)^2 + \gamma^2} \ .
\end{equation}
Multiplying \eqref{shear_conteqrho+} with $g(\omega,\eta)$ and integrating over $(\omega,\eta)$, we can use Fubini's theorem (on the assumption of integrability of $\Omega g \hat \rho_+$ and $H g \hat \rho_+$) and compute the double integral by changing the order of integration.
First, we can evaluate the integral over $\omega$ by applying the residue theorem as in Section \ref{PDS} and then move on to the second integral, which reads
\begin{align*}
\partial_t \hat \rho_+&(\theta, \omega_0-i\delta,-i\gamma,t) = \\
- \int_{-\infty}^\infty \partial_\theta & \left\lbrace \left[ \Omega(\omega_0-i\delta, \eta, t) + \frac{1}{2i} \left( H(\eta,t) \mathrm{e}^{-i\theta} - H(\eta,t)^\ast \mathrm{e}^{i\theta} \right) \right] g_2(\eta) \hat \rho_+(\theta, \omega_0-i\delta,\eta,t) \right\rbrace d\eta \ .
\end{align*}
While the term $\int \Omega g_2 \hat \rho_+$ can be evaluated at the pole $\eta = \eta_0 \pm i \gamma$ ($\pm$ depending on the contour of integration, which again depends on the coupling $K$, see also Ref.~\onlinecite{MontbrioPazoShear2011}) , we have to assure that the product $H(\eta, t) g_2(\eta)$ vanishes for $\mathrm{Im} (\eta)  \to \pm\infty$.
Indeed, the linear growth of $H$ in $\eta$, see \eqref{shear_prop}, will be dominated by the exponential decay of $g_2$, such that the residue theorem can be applied here, too, which results finally in \eqref{eq17}\&\eqref{eq18}, from which the claim follows as presented in Section \ref{PDS}.
As has been shown in Section \ref{general}, the restrictions to unimodal Lorentzians can be dropped and the OA attractiveness is sustained.
Here we can even handle $\delta$-functions as long as one of the partial distribution functions has finite width: due to the special form of $\Omega(\omega, \eta, t)$, the OA ansatz holds for homogeneous frequencies $\omega_j = \omega$ while the shear is heterogeneous and the coupling $K > 0$ does not vanish.

The case in which the joint distribution $g(\omega, \eta)$ is no longer uncorrelated, i.e. if the first equality in \eqref{shear_uncorr} fails, demands a more careful investigation in order to estimate the long-time evolution of $\hat \rho_+$.
Although the ultimate goal is to categorize adequate joint distributions that allow for the OA ansatz, there might appear a variety of uncertainties for a general proof.
For instance, to the best of our knowledge it is an open problem whether and how singularities can appear in joint distributions given smooth marginal distributions.
This issue becomes even more intricate in the case for multi-dimensional parameters $\eta \in \mathbb{R}^n, \ n \in \mathbb{N}$.
However, there are certain approaches using the OA ansatz for parameter-dependent systems with correlated joint distributions, which we would like to briefly revise.

The introduction of shear into the oscillator system shows how an additional parameter can be treated as a random variable and thereby changing the natural frequency and driving field of the original Kuramoto model.
A more fundamental approach has been presented by Petkoski and co-workers in Refs.~\onlinecite{PetkoskiPRE2012,PetkoskiPRE2013,Iatsenko2013,Iatsenko2014}:
Given the Kuramoto model with heterogeneous natural frequencies, they assume the coupling strengths to be drawn from a distribution function.
That is, their model reads
\begin{equation}
\label{Kuramoto_Spase}
\dot \theta_j = \omega_j + \frac{K_j}{N} \sum_{k=1}^N \sin(\theta_k - \theta_j) 
\end{equation}
with $(\omega, K)$ following a joint distribution $g(\omega, K)$.
Given the strong resemblance between their numerical simulations and the predictions via the OA ansatz, the authors realized that the latter ``formulas were derived on the assumption of at least asymptotic validity of the OA ansatz.'' \citep{Iatsenko2013}
They also investigated necessary initial conditions with respect to their analytic continuation and applicability to the OA ansatz.
Unfortunately, they did not prove this their system dynamics \eqref{Kuramoto_Spase} does not belong the classes of systems considered in the proofs by Ott and Antonsen\cite{OttAntonsen2008,OttAntonsen2009,OttAntonsenComment}.
Recall, a general characterization of correlated joint distribution $g(\omega,K)\neq g_1(\omega) g_2(K)$ that are applicable for the extended OA ansatz is hardly feasible.
However, for three examples used in literature we can prove that the OA manifold defines the asymptotic evolution of the whole system.

First, let $g(\omega, K) \sim \delta(K-k) \left[ \omega^2 + \mathrm{e}^{-\omega^2}\right]^{-1}$, see Fig.1 in Ref.~\onlinecite{Iatsenko2014}.
The specific form with the $\delta$-function in $K$ reduces system \eqref{Kuramoto_Spase} to the common Kuramoto model with heterogeneous frequencies $\omega \propto g_1(\omega) = \left[ \omega^2 + \mathrm{e}^{-\omega^2}\right]^{-1}$, which can be dealt with along the proof of the original OA ansatz.

The other two examples are more elaborate in that the joint distribution functions are given by\cite{Iatsenko2013}
\begin{equation}
g(\omega,K )  = (1-p)\delta(K-K_1) L(\omega; \omega_0, \gamma_1)+ p\delta(K-K_2) L(\omega; - \omega_0, \gamma_2) \ ,  \label{jdone}
\end{equation}
with $p \in \left( 0,1 \right]$, and
\begin{equation}
g(\omega,K )  = \Gamma(K) \sum_{n=1}^{N_q} q_n L(\omega ; \omega_n, \gamma_n) \ , \text{with } \quad \sum_{n=1}^{N_q} q_n(K) = 1 \ . \label{jdtwo}
\end{equation}
Here, $L(\omega ; \omega_n, \gamma_n)$ denotes a Lorentzian of width $\gamma_n >0$ and centered around $\omega = \omega_n$, and $\Gamma(K)$ is a multimodal-$\delta$-function.
For properly chosen $q_{1,2}$ \eqref{jdtwo} can be regarded a generalization of \eqref{jdone} so that it is sufficient to deal with the former.
For simplicity, let us consider $N_q = 2$, i.e. $g(\omega,K)$ to be a bimodal joint distribution. 
Employing $g(\omega,K)$ in the definition of the order parameter \eqref{orderparam2}, we see that we can decompose it into $z(t) = q_1  z_1(t) + q_2 z_2(t) $ with $q_1 + q_2 = 1$.
That is, we can view our system as two all-to-all coupled populations with population-specific coupling strengths $K_{1,2}$.
Given that the frequency distributions are Lorentzians of finite width $\gamma_{1,2}$ we can apply the results for two-population/bimodal Kuramoto models as in Refs.~\onlinecite{MartensExactResults2009,Pietras2016}, which confirms the attractiveness of the OA manifold for this kind of joint distributions.
The case of multiple Kuramoto populations with specific coupling strengths can be approached by transforming the system into one global system whose oscillators' frequencies follow a multimodal distribution consisting of weighted inhomogeneous unimodal distributions, which can mirror the underlying coupling topology across populations\cite{PietrasInProgress}.

It is true that the examples mentioned above are not exhaustive but rather represent a small set of a broad variety of joint distribution functions. 
Nevertheless, we believe that our results may be a major breakthrough for the applicability of the OA ansatz for systems with more intricate distribution functions.

\subsection{Heterogeneous mean field models}
While the general case of uncorrelated joint distributions has already been covered in the preceding Section \ref{subsec_shear}, we would like to concentrate on the specific derivation of the heterogeneous mean field model.
Recall the standard Kuramoto model on a given network,
\begin{equation}
\dot\theta_j = \omega_j + K \sum_{k=1}^N a_{jk} \sin(\theta_k - \theta_j) \ ,
\end{equation}
where $K$ is the coupling strength and the adjacency matrix is given by $A = (a_{ij})_{i,j=1,\dots,N}$.
We substitute the adjacency values $a_{jk} \in \{ 0,1 \} $ by their expectation values $\left\langle a_{jk} \right\rangle \in \left[ 0,1 \right]$, which are given by
\begin{equation}
\label{HMF_expect}
\left\langle a_{jk} \right\rangle = \frac{\eta_j \eta_k}{N \left\langle \eta \right\rangle } \ .
\end{equation}
Introducing the complex order parameter as 
\begin{align*}
z =  \frac{1}{N\left\langle \eta \right\rangle} \sum_{k=1}^N \eta_k \mathrm{e}^{i\theta_k} \ ,
\end{align*}
the dynamics for all nodes with the same degree $\eta_k$ read
\begin{align*}
\dot\theta_k = \omega_k + K \eta_k \mathrm{Im} ( z \mathrm{e}^{ - \theta_k}) \ .
\end{align*}
In this special form, in which the single nodes are replaced by block-degree variables, we returned to the all-to-all coupling.
For a given degree distribution $P(\eta)$ property \eqref{HMF_expect} also holds in the continuum limit $N \to \infty$ where the governing dynamics read
\begin{equation}
\label{HMF_velo}
\partial_t \theta(\eta,t)  =  \omega + \mathrm{Im} \left[ K \eta z(t) \mathrm{e}^{-i\theta} \right] \ ,
\end{equation}
with $\omega$ and $\eta$ being drawn from a joint distribution $g(\omega, \eta) = P(\eta) g_1(\omega)$.
As before we can introduce a phase distribution function $\rho(\theta, \omega, \eta, t)$, which fulfills the continuity equation $\partial_t \rho + \partial_\theta (v \rho) = 0$ with $v$ the right-hand side of \eqref{HMF_velo}.
Note, however, that depending on the underlying network topology and its degree distribution $P(\eta)$, one has to choose the domain of $\eta$ properly.
In the case of a scale-free network, the degree distribution follows $P(\eta) \propto \eta^{-\gamma}$ with $\gamma > 1$.
Hence the normalization conditions for the distribution function $\rho$ obey
\begin{align*}
\int_1^\infty \int_0^{2\pi} \rho(\theta, \omega, \eta, t) \ d\theta d\eta = g_1(\omega) \ \ \text{and} \ \
\int_{-\infty}^\infty \int_0^{2\pi} \rho(\theta, \omega, \eta, t) \ d\theta d\omega = P(\eta) \ .
\end{align*}
We can apply the OA ansatz as before.
By the same reasoning as in Section \ref{subsec_shear}, we can so prove the OA attractiveness for heterogeneous mean field models, rendering also non-globally coupled oscillator networks applicable to have their mean field dynamics evolved on a low-dimensional manifold.

\subsection{Non-local coupling}
Two months before Ott and Antonsen published their ansatz, Ko and Ermentrout investigated the creation of partially locked states in a network of identical all-to-all coupled oscillators due to inhomogeneous coupling\cite{KoErmentrout2008}.
Instead of heterogeneity of the oscillators' frequencies, it was the coupling heterogeneity that led to partial synchronization.
Carlo Laing analytically investigated this network of globally coupled oscillators with coupling strengths drawn from a power-law distribution\cite{laing2009dynamics} along the line of the OA ansatz -- recall the resemblance to the heterogeneous mean field approach for scale-free networks.
Assuming ``nearly'' identical oscillators, i.e. the frequencies $\omega$ were drawn from a Lorentzian with width $0 < \Delta \ll 1$, he could verify the earlier results that were derived via a self-consistency argument\cite{KoErmentrout2008}, and extend them by including a thorough bifurcation analysis.
Our findings in Section \ref{subsec_shear} put these results on a solid mathematical ground.

Of particular interest is Laing's work on a ring of oscillators\cite{laing2009chimera,laing2009dynamics}.
For a given ring topology, the typical coupling scheme is neither local neighbor-to-neighbor, nor global coupling.
Instead, the oscillators are non-locally coupled via a coupling kernel $G$.
We assume that each oscillator $k = 1,\dots, N$ has some fixed spatial position $x_k \in [-\pi, \pi]$, a natural frequency $\omega_k$ drawn from a continuous distribution function $g(\omega)$ with non-zero width, and interacts with the others depending on the distance between their sites modulo periodic boundary conditions.
The governing dynamics read
\begin{equation}
\dot \theta_k = \omega_k - \frac{2\pi}{N} \sum_{j=1}^N G(x_k - x_j) \sin (\theta_k - \theta_j + \alpha) \ ,
\end{equation}
where $\alpha$ is a phase-lag parameter and $G \colon \mathbb{R} \to \mathbb{R}$ a continuous even and $2\pi$-periodic coupling function\cite{omel2013coherence}.
We retrieve global coupling, if $G\neq 0$ is constant.
Commonly used coupling functions $G$ are of exponential form $G(x) \sim \mathrm{e}^{-\kappa |x|}$ with $\kappa > 0$, or of trigonometric form $G(x) = 1/2\pi (1 + A \cos x + B \sin x)$ with $A > 0, B \geq 0$.
The reflection symmetry of $G$ is lost for $B \neq 0$.
In the continuum limit, the velocity field \eqref{velofieldnew} becomes
\begin{equation}
\begin{gathered}
\partial_t \theta = \omega + \mathrm{Im} \left[ H(x,t) \mathrm{e}^{-i\theta} \right] \ , \\
H(x,t) \mathrm{e}^{i\alpha} = \int_{-\pi}^\pi G(x-y) \int_{-\infty}^\infty \int_0^{2\pi} \rho(\theta, y,\omega,t)\mathrm{e}^{i\theta} \ d\theta d\omega dy \ .
\end{gathered}
\label{nonlocal_coupling}
\end{equation}
While the inner two integrals have the form of a local complex order parameter $z(y,t)$, measuring the synchronization degree of oscillators around $y$, we can interpret the last integral as a convolution of the local order parameter with the (spatial coupling) kernel $G$.
In particular, we can regard the dynamics $\partial_t \theta(x,t)$ of an oscillator at position $x$ as being controlled by the local mean field $H(x,t)$.
Unlike the case of global coupling, the order parameter has become space-dependent and thus the driving field.
However, a similar ``physical picture'' as for global coupling is valid:
practically we deal with an assembly of independent oscillators under the control of a common forcing field\cite{KuramotoBatto2002,omel2013coherence}.
We now go a step further and interpret the space variable $x$ as a subpopulation index\cite{wolfrum2011spectral}.
Equivalent to the block-degree variables in the heterogeneous mean field approach, we consider the subpopulation index as a parameter that follows a particular, in this case a uniform, distribution function.
Hence, \eqref{nonlocal_coupling} represents the governing dynamics of a parameter-dependent system, for which we proved the OA attractiveness in the preceding sections.

\section{Relaxation dynamics}
\label{relaxdynamics}

As discussed, we allow time-varying parameters to affect the oscillator dynamics.
The change of parameters comes with its time scale(s).
The change can be periodic.
This periodicity may also influence the evolution of the mean field and thereby the OA manifold.
Therefore, the relation between this periodicity and the characteristic time of the system to approach the manifold needs to be investigated.
If the relaxation dynamics onto the manifold is way slower than the characteristic time scale of the time-varying manifold itself, then our findings will remain true for the limit $t \to \infty$.
They are, however, of minor interest for describing the transient behavior of the mean field.
Several numerical results\cite{MontbrioPazoRoxin2015,PetkoskiPRE2012,Iatsenko2013,Iatsenko2014} suggest that the relaxation to the OA manifold is reasonably fast, in some cases even instantaneous.
To address this analytically, we briefly recall the proof for the attractiveness from Section \ref{PDS}.
After having Fourier expanded the phase distribution function $\rho(\theta, \eta, t)$, and then decomposed the positive Fourier modes into a part that already lies on the manifold, $\hat \rho'_+$, and a residual part $\hat \rho_+$, we showed how the latter converged to zero in a weak sense, cf. \eqref{Claim}.
We can extract the relaxation time to the OA manifold from out of the proof:
From \eqref{eq17}\&\eqref{eq18} we obtain a solution $f_+(\theta, t) = \hat \rho'_+(\theta, -i \sigma, t)$, with $\sigma' > \sigma > 0$ where $\hat \rho'_+(\theta, \eta, t)$ admits an analytic continuation into the strip $S = \{ \eta \in \mathbb{C} \ | \ -\infty \leq \mathrm{Re} (\eta) \leq \infty \ , \ 0 \geq \mathrm{Im}(\eta) \geq -  \sigma' \} $;
the solution \eqref{charsol} obeys
\begin{align*}
\tilde f_+(w, t) = \tilde f_+(W(w,0), 0) \exp \left[ -\mu (w,t) \right] \ , 
\end{align*}
hence the relaxation time $\tau$ is by definition
\begin{equation}
\label{relax_time}
const \cdot \exp(-t/\tau) =  \exp \left[ -\mu (w,t) \right] \quad \Rightarrow \quad \tau =  \frac{t}{\mathrm{Re} \left[\mu(w,t)\right] } \ .
\end{equation}
Put differently, $\mathrm{Re} \left[\mu(w,t)\right]$ scales with $\sigma t$, such that $\tau = 1/ \sigma$.
The wider the frequency distribution becomes, the larger $\sigma$ can be chosen.
Thus, one may argue that the characteristic time scale decreases with increasing heterogeneity among the single oscillators.
This relation has already been noted for a particular example of a Lorentzian frequency distribution by Ott and Antonsen in Ref.~\onlinecite{OttAntonsen2008}. It has been investigated in more detail by Petkoski and Stefanovska for the non-autonomous Kuramoto model\cite{PetkoskiPRE2012}.
Interestingly, there is an intrinsic relation between the frequency inhomogeneity and the coupling strength.
Therefore, at critical coupling strengths, which distinguish different dynamical regimes, the relaxation times tend to infinity, which has been reported independently by Petkoski et al.\cite{PetkoskiPRE2012} and Yoon et al.\cite{Yoon2015} for the full Kuramoto network, its non-autonomous version and the heterogeneous mean field model.

For the non-autonomous case we would like to mention that the proof presented in Section \ref{PDS} entirely holds for continuous time-varying parameters.
Introducing discontinuities in either the frequency $\Omega$ and/or the driving field $H$, however, will eventually lead to a non-continuous right-hand side of \eqref{eq18} -- due to $H$ itself, or via the order parameter $z$, which absorbs the time-varying part of $\Omega$ and influences $H$ directly or indirectly.
While employing the method of characteristics still can be performed, estimating the integral in \eqref{charhelp} cannot exploit the continuity assumption and a proper evaluation has to be circumvented.
In spite of this sinister outlook, numerical results remain promising; for instance, the simulations in Ref.~\onlinecite{MontbrioPazoRoxin2015} with a square input function (Fig.2a,c,e,g).
A possible way to overcome this obstacle might be to approximate the jumps by smooth sigmoid functions, which might be valid as long as the height of the jumps is lower than their length.
Another more rigorous approach might be to find weak solutions for (\ref{eq17}~\&~\ref{eq18}) and estimate their long-time behavior.
There, a starting point could be the very recent results by Dietert, Fernandez and co-workers, who investigated stability properties of different dynamical regimes of the Kuramoto model in a mathematically rigorous way, confirming the exponential decay to the manifold\cite{Dietert2016, Fernandez2016, DietertFernandez2016}.
More details are way beyond the scope of our paper.

Interestingly, the approach by Dietert and others is based on the idea of ``Landau damping'' in plasma physics.
Strogatz, Mirollo and Matthews were the first who incorporated this concept in order to understand relaxation dynamics of the Kuramoto model\cite{StrogatzMiMa1992,Strogatz2000}.
They showed that for frequency distributions $g(\omega)$ supported on the whole real axis, the decay towards the incoherent state is exponentially fast for coupling strengths below the critical threshold, $K < K_c$.
If $g(\omega)$ has compact support, i.e. $g$ is non-zero only on a compact interval $\left[ -\gamma, \gamma \right] \subset \mathbb{R}$, $0 < \gamma < \infty$, the rate may be considerably slower, even polynomial.
In the example they used to illustrate their result, the authors assumed the frequencies $\omega$ to be distributed uniformly on $\mathcal{I} = \left[ -\gamma, \gamma \right] $, i.e. $g(\omega) = 1 / 2\gamma$ if $\omega \in \mathcal{I}$, and $0$ otherwise.
The jump discontinuities of $g$ on $\partial \mathcal{I}$, however, prohibited an analytic continuation of $g$ into a strip $S$ in the lower complex $\omega$-plane, contradicting the required conditions for applying the OA ansatz\cite{OttAntonsenComment}.
That is why the proofs above cannot be applied here, and our argumentation about the relaxation times remains unaffected.

Last but not least, we would like to add that decay times typically depend on initial conditions.
Pikovsky and Rosenblum pointed out that for identical macroscopic, i.e. mean field, initial conditions the microscopic initial states can lead to very different transient dynamics towards the OA manifold, see Section 3.2 in Ref.~\onlinecite{PikovskyRosenblum2011}.
A more thorough investigation about this specific topic has not been undergone, yet, but might shed light on the underlying dynamics of the microscopic variables of large oscillatory systems in contrast to its mean field behavior.

\section{Discussion and conclusion}
The OA ansatz has proven rather fruitful for investigating the macroscopic behavior of systems of coupled phase oscillators in terms of a low-dimensional system.
Although parameter dependence has already been mentioned in Ott and Antonsen's original work, parameters were merely considered auxiliary variables and the velocity field was required to incorporate the phase only through a sinusoidal coupling term.

Our main result was to prove that the $\eta$-dependence sustains the time-asymptotic attractiveness of the OA manifold for systems of coupled oscillators.
For this we required that the driving field $H$ does not have singularities in the complex $\eta$-plane and that it diverges at most sub-exponentially for $\mathrm{Im} (\eta) \to -\infty$, next to the conditions in the original Ott \& Antonsen formulation \cite{OttAntonsen2008,OttAntonsen2009}. Furthermore, we assumed the frequency $\omega(\eta, t)$ to be linear in $\eta$.
We were able to depict the proof step by step.
Subsequently we loosened the restrictive assumptions and showed that our results remain valid for a much broader class of distribution functions $g(\eta)$ as well as more complex dependencies of the driving field $H(\eta)$ and the natural frequencies $\omega(\eta)$ on the parameter $\eta$.

Although the main idea of introducing a common parameter $\eta$ was to correlate the driving field and the natural frequency with their specific oscillator, our proof is identical for the case when $\eta$ does only influence the mean field dynamics.
By this, we have proved the claim in Ref.~\onlinecite{OttAntonsenComment} that the OA manifold remains attractive in the ``weak'' parameter-dependent case when $H$ depends on ``other non-phase-oscillator variables obeying auxiliary dynamical systems.''

Common choices of $H$ and $\omega$ usually fulfill the aforementioned assumptions as stated in Section \ref{PDS}.
That is, our result can be immediately applied in a variety of circumstances.
Here, we highlighted an application in mathematical neuroscience.
By this, our findings strengthen the theory of coupled theta neurons:
The many recent numerical findings of Ref.~\onlinecite{MontbrioPazoRoxin2015} and the references therein are finally set in a solid mathematical framework.
Moreover, the link between QIF neurons and theta neurons has been underscored by proving the attractiveness of the Lorentzian ansatz. 

We generalized and extended existing proofs for non-autonomous systems. In particular we addressed the Winfree model, which is biologically more realistic than the Kuramoto model and therefore closer to applications.
We also addressed coupled oscillatory systems with an additional shear parameter, another important tool to render the Kuramoto model more realistic.
The major novelty was our rigorous proof of the OA attractiveness for systems with uncorrelated joint distribution functions when more parameters than only the natural frequencies are treated as a random variable.
This finding opened the way for networks with specific underlying coupling topologies  other than the restrictive global coupling.
Using the heterogeneous mean field approach, we showed how these networks can be treated along the OA ansatz.
First steps were also taken in the direction of correlated joint distributions.

All in all, we consider the explicit dependence on an additional parameter $\eta$ of both the oscillator's phase and the (non-sinusoidal) components an important extension introducing an intrinsic relation between phase, frequency, and driving field of an oscillator.
The latter two are correlated with the phase so that the $\eta$-dependence does not allow for applying the original theory.

Still, there are several open problems concerning the mean field dynamics of an oscillatory system and its description by a low-dimensional system.
A first urgent one is the case of $\delta$-peaked frequency distributions.
Numerical simulations\cite{omel2014partially} and heuristic arguments hint at convergence of the OA manifold, where a proper mathematical derivation is omitted under the pretence of ``nearly identical oscillators''\cite{MartensBickPanaggio2016,laing2009chimera, laing2012disorder}.
A thorough proof would render the OA ansatz rigorously applicable to ``chimera states'', a topic that is particularly en vogue; see, e.g., the recent review paper by Panaggio and Abramscite\cite{panaggio2015chimera}.
Importantly, such a proof has to circumvent the main argument of Ott and Antonsen's original proof, where the width $\Delta > 0$ of the distribution $g(\omega)$ allowed for a consequent evaluation of the mean field dynamics.
On the other hand, Pikovsky and Rosenblum\cite{PikovskyRosenblum2008} already showed that more complicated dynamics can emerge from the OA manifold when describing the system along the Watanabe-Strogatz (WS) ansatz\cite{WatanabeStrogatz1994}.
Deviations from the OA ansatz appear only if the WS constants of motion are not uniformly distributed over the whole domain, but only over a compact subset.
Given (a) the direct correspondence between the constants of motion and the initial conditions of phases in the OA ansatz\cite{PikovskyRosenblum2011,WatanabeStrogatz1994}, and (b) the necessary requirements on (analytic continuation properties of) the initial conditions, it may be worth investigating the influence of nonuniform distributions of the constants of motion and whether this may hinder the initial conditions of phases to satisfy the requirements of the OA ansatz.

Another intriguing open problem is whether the mean field dynamics is attracted by a low-dimensional manifold when the parameter-dependence of the frequency and driving field is extended by an explicit dependence on the phase.
A recent example is given by Laing\cite{laing2015chimeras}, who considered the driving field $H$ to follow a dynamics that explicitly depends on the phase $\theta$.
This system exhibits partial synchronization patterns, which are also covered by the OA ansatz, but any attempt to apply the OA ansatz has been avoided ``due to the dynamics of the extra variables.''\cite{laing2015chimeras}

When the coupling term incoporates higher harmonics, see, e.g., Refs.~\onlinecite{SkardalOttRestrepo2011,Terada2016}, no low-dimensional analytic solution for the mean field evolution has been found.
This is another open question whether further generalizations of the work of Ott and Antonsen [\onlinecite{OttAntonsen2008}] can be rigorously manifested.
We believe that our current proof for parameter-dependent networks is a good starting point for tackling these important issues.

\begin{acknowledgements}
This project has received funding from the European Union's Horizon 2020 research and innovation program under the Marie Sk{\l}odowska-Curie grant agreement \#642563 (COSMOS).
We would like to thank Ernest Montbri\'{o}, Oleh Omel'chenko, Spase Petkoski and Arkady Pikovsky for many fruitful discussions.
\end{acknowledgements}

%\bibliographystyle{plain}
%\nocite{*}

%\bibliography{refcomment}

 \newcommand{\noop}[1]{}

\end{document}